\documentclass[galaxies,review,accept,moreauthors,pdftex]{Definitions/mdpi}

\graphicspath{{figures/}}
\usepackage{xcolor}

\firstpage{1} 
\pubvolume{1}
\issuenum{1}
\articlenumber{0}
\pubyear{2022}
\copyrightyear{2022}
\externaleditor{Academic Editor: {Junhui Fan}} 
\datereceived{}
\dateaccepted{} 
\datepublished{} 
\hreflink{https://doi.org/} 

\usepackage[labelformat=simple]{subfig}

\Title{{Blazars} at Very High Energies: Emission Modelling}

\TitleCitation{Blazars at Very High Energies: Emission Modelling}

\Author{Hélène Sol
 * and Andreas Zech \orcidA{} *}

\AuthorNames{Hélène Sol and Andreas Zech}

\AuthorCitation{Sol, H.; Zech, A.}

\address[1]{%
Laboratoire Univers et Th\'eories, Observatoire de Paris, Universit\'e PSL, CNRS, Universit\'e Paris Cit\'e,  5 Place Jules Janssen, F-92190 Meudon, France
}

\corres{\hangafter=1 \hangindent=1.05em \hspace{-0.82em} 
Correspondence: helene.sol@observatoiredeparis.psl.eu (H.S.);  andreas.zech@observatoiredeparis.psl.eu (A.Z.)
}

\abstract{Blazars are very broadband cosmic sources with spectra spanning over twenty orders of magnitude in frequency, down to the 100 MHz regime in the radio range, up to VHE at several tens of TeV. The modelling of their spectral energy distribution at high energies currently considers two main classes of models, leptonic and lepto-hadronic, which both succeed fairly well in describing the observed spectra for the two populations of blazars, namely BL Lac objects (BL Lacs) and  flat spectrum radio quasars (FSRQs). However they are both confronted with difficulties, in particular to reproduce flaring phenomena monitored with a good multi-spectral and temporal coverage, or to reproduce extreme sources which challenge the basic descriptions. Such a situation has led to a diversity of specific scenarios, the positioning of which in relation to the general context of the sources is generally not clearly fixed. The identification of the dominant particle acceleration mechanism at work and a better understanding of the location of the TeV emitting zone would make it possible to break the degeneracies between models. Multi-wavelength and multi-messenger studies should also help in this regard, with the perspective to elaborate a general reference scenario 
of blazars and~AGNs. }


\begin{document}

\section{Introduction}

The broadband spectral energy distribution (SED) from blazars can stretch over the full observationally accessible spectrum, from~the radio-band up to very-high-energy (VHE) gamma-rays. This continuum emission
is dominated by non-thermal radiation in the form of two spectral bumps, with~the first peaking in the optical, ultraviolet or X-ray range and the second in the high-energy or VHE gamma-ray band. Depending on the source, non-thermal emission from an extended radio jet and thermal emission from the host galaxy, the~accretion disk and the dusty torus can lead to a more complex~SED. 

The comparison of multi-wavelength (MWL) emission from blazars with numerical models that calculate radiative emission and transfer for given assumptions on the particle content and characteristics of the emission regions, represents the principal tool used to probe the microphysics inside the source and its physical parameters. Information from the VHE part of the SED is a necessity to constrain model parameters for high-frequency peaked (BL Lac) objects, which radiate a significant part of their overall emission in that energy range, but~observational VHE constraints can also be important for softer sources, since they probe the very limit of particle acceleration and are also most sensitive to photon-photon absorption on internal or external photon~fields.

VHE observations are also crucial for the characterisation of the variability time scales involved in variable emission. The~short cooling time at these energies leads to the most rapid variability, imposing rigorous constraints on the size of the emission region, due to light travel time~arguments.  
 
We first provide a short review of the main emission scenarios, their applications to different blazar types and their limitations in Section~\ref{sec:models}. Then we turn our attention to the underlying particle acceleration mechanisms in Section~\ref{sec:acceleration}. A~critical question, which might not have a unique answer, concerns the location of the emission zone(s) responsible for the VHE emission (Section~\ref{sec:zone}). We finally discuss how MWL and multi-messenger (MM) observations constrain the large variety of proposed emission models in Section~\ref{sec:mwl} and conclude in Section~\ref{sec:conclusion} with a brief outlook on the spin-offs of a better understanding of blazar~physics.


\section{Blazar Emission~Modelling}
\label{sec:models}

While spectra observed in a limited wavelength range can be directly fitted with power-law and log-parabola functions or simple physical models, the~interpretation of the full SED is usually based on a comparison with the synthetic MWL emission from radiative models of different degrees of complexity. Such modelling of the SED can exploit proper fitting in the case of simple models (e.g.,~\cite{Nigro2022}), but~for models with a high number of free parameters and for the often sparse and not strictly simultaneous MWL data sets, an~adjustment ``by eye'' is still the standard approach. Given the limited observational coverage and the intrinsic degeneracy of even the simplest emission models, the~objective of SED modelling is rather to identify a range of physically acceptable models that can explain the data, than~to find the best formal fit model, which might spuriously lead to a set of extreme parameter values. 

\subsection{Leptonic Emission~Modelling}
While the low-energy bump in the non-thermal emission is generally ascribed to synchrotron emission from electrons or pairs inside the jet, the~high-energy bump can be interpreted with leptonic or hadronic processes. 
In the standard leptonic approach, electrons or pairs up-scatter the synchrotron radiation they have produced and photons from external fields to (very) high energies through the Inverse Compton process (\mbox{e.g.,~\cite{Konigl1981, Dermer1993, Sikora1994, Levinson1995, Inoue1996, Ghisellini1996, Ghisellini1989, Katarzynski2001}). }

If external photon fields can be neglected, the~stationary single-zone synchrotron-self-Compton (SSC) model provides the most basic approach to describe the steady MWL emission (e.g.,~\cite{Bednarek1997, Tavecchio1998}). For~a phenomenological approach, the~emission region is usually described as a spherical ``blob'' filled with a tangled magnetic field of homogeneous strength and a single particle distribution parameterised by a broken power law or a logparabola, which generally provide the necessary free parameters to represent the observed shape of the synchrotron and inverse-Compton bumps. Alternatively, a~particle spectrum is injected and evolved, based on a differential (Fokker--Planck) equation, which can account for particle injection, acceleration, radiative and adiabatic cooling, and~particle escape, into~a steady-state particle spectrum. 
 A general form is given by ~\cite{kardashev1962, tramacere2011} as:  
\begin{equation}
\begin{array}{c}
 		\dfrac{\partial N_\text{e}(\gamma,t)}{\partial t} = \dfrac{\partial}{\partial \gamma} \left[ (b_\text{c} \gamma^2 - a\gamma - 2D_0\gamma ) \cdot N_\text{e}(\gamma,t) \right] \, +  \\ \\
		 \dfrac{\partial}{\partial \gamma}\left(D_0\gamma^2 \dfrac{\partial N_\text{e}(\gamma,t)}{\partial \gamma} \right) - \dfrac{N_\text{e}(\gamma,t)}{t_{\text{esc}}} + Q_{\text{inj}}(\gamma,t)
 		\label{equ:fokker_planck}
 		\end{array}
 		\end{equation}
Here 
 $b_c \gamma^2$ is the radiative cooling rate, $a \gamma$ the Fermi-I acceleration rate, $D_0 \gamma^2$ the energy diffusion coefficient, $t_{esc}$ the particle escape time and $Q_{inj}$ the injection function. This particular description is appropriate in the ``hard-sphere'' scattering~approximation.		

The whole emission region is moving towards the observer at relativistic speed, with~bulk Lorentz factor $\Gamma$  under a small viewing angle $\theta$, leading
to a Doppler boosting of the emission that is characterized by the bulk Doppler factor 
$\delta = 1 / \Gamma(1 - \beta \cos\theta)$ 
Adiabatic expansion of the ``blob'' during the considered duration of its emission is mostly neglected in such models. Under~certain conditions, the~emission region can also be interpreted as a standing shock through which plasma is moving at relativistic bulk speed. The~single-zone SSC scenario 
requires only 8 or 9 free parameters and,
even with this simplistic approach, 
usually succeeds in reproducing the available data for most blazars deteced at VHE, but~does not provide a unique solution for a given MWL SED. It is however possible to carry out a scan of the available parameter space and to identify the acceptable range of solutions for a given confidence interval~\cite{Tavecchio1998, Cerruti2013}.

\begin{figure}[H] 
\includegraphics[width=13 cm]{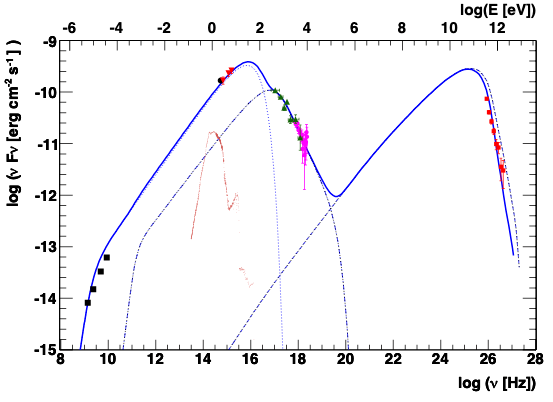}
\caption{A typical  
 stationary SSC model for the high-frequency peaked BL Lac object PKS 2155$-$304. See text for details. Credit: \cite{hess2012}, A\&A, reproduced with permission @ESO.\label{fig:ssc}}
\end{figure}

Figure~\ref{fig:ssc} shows the interpretation of an SED of a high-frequency peaked BL Lac object, where the synchrotron and SSC components of a compact emission region account for the X-ray and VHE emission, while additional 
synchrotron emission from the extended jet dominates the lower frequencies. Such an extended component is generally required to correctly describe at least the low-frequency radio emission. In~this case, emission from the host galaxy (thin brown line) is completely dominated by the non-thermal emission. Absorption by the extragalactic background light (EBL) is noticeable at VHE although it remains weak at small~redshifts.

Leptonic models that account for interactions with external photon fields, from~the accretion disk, broad line region, dusty torus, surrounding starlight (or photons from the EBL and the CMB in the case of extended jets or intergalactic particle cascades), become rapidly more complex, even when using a single-zone approach. The~large number of additional free parameters to describe these photon fields can be reduced either through direct observational constraints or by using scaling relations that link the different contributions to the accretion disk luminosity or 
supposed mass accretion rate and radiative efficiency of the central engine~\cite{Ghisellini2009}. The~inclusion of external Inverse Compton (EIC) radiation makes it possible to reproduce the high Compton dominance observed in luminous blazar classes (e.g.,~\cite{Sikora1994}).

\begin{figure}[H]
\includegraphics[width=13 cm]{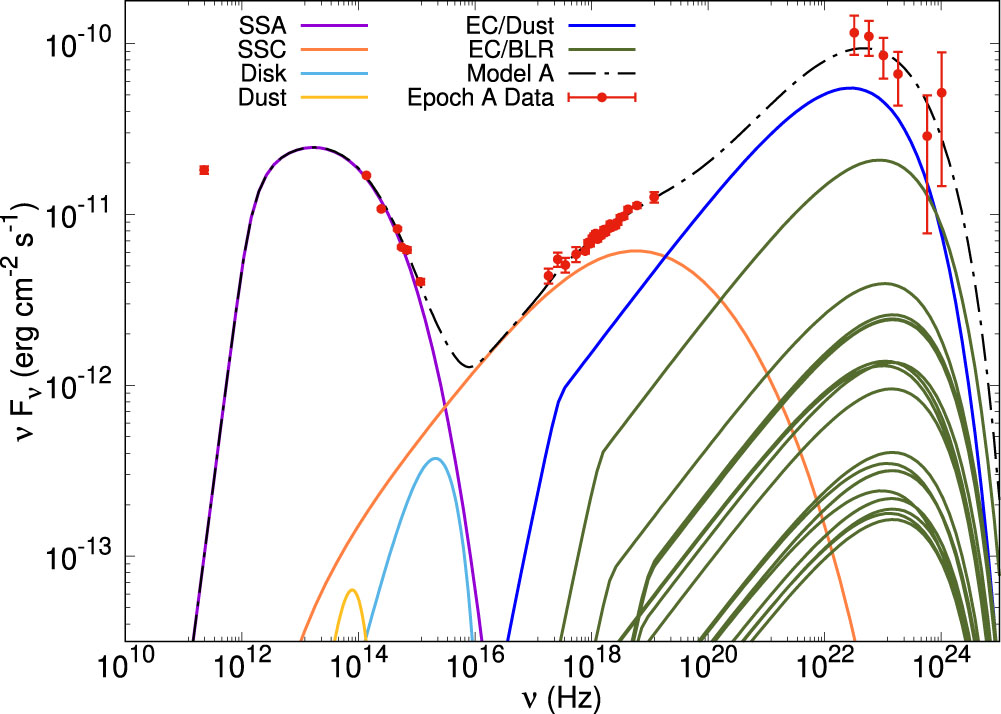}
\caption{A typical stationary model, including a range of EIC components, for~the FSRQ 3C\,279 (taken from~\cite{lewis2019} based on a code by~\cite{Finke2016}). Credit: \cite{lewis2019} @AAS. Reproduced with~permission. \label{fig:eic}}
\end{figure}

Figure~\ref{fig:eic} shows a stationary model including, in~addition to synchrotron and SSC emission, EIC radiation from several external photon fields, while the SSC emission dominates at X-ray energies, the~high Compton dominance in the gamma-ray range is due to these latter components. Direct emission from the accretion disk and the dust torus is strongly dominated by the non-thermal emission in this particular~case.

While such steady-state models can be used to broadly characterize different activity states of blazars, in~order to truly model rapidly varying emission, as~encountered during blazar flares, a~time-dependent description of the electron distribution and/or of the source parameters is required.
Flares can then be modelled through an interplay of particle injection or acceleration with particle cooling and escape, following, e.g., Equation~(\ref{equ:fokker_planck}) as a function of time, or~through sudden changes in the magnetic field, Doppler factor or physical extension of the emission region (e.g.,~\cite{Joshi2011, Bottcher2019, Dmytriiev2021}). 

The radiative processes involved in leptonic models, synchrotron and Inverse Compton radiation, as~well as photon absorption through pair production, are well known and documented. Beyond~the single-zone scenario, a~description of the emission from the extended jet is usually achieved by representing the latter as a combination of homogeneous slices or cells and by calculating the radiative transfer along a given line of sight, while this is straightforward for the synchrotron and EIC emission, the~calculation of SSC emission from such an extended region is more complicated, since one ideally needs to take into account the radiation emanating from all regions of the jet beyond the SSC emission region under study~\cite{Katarzynski2003,Graff2008}.  

\subsection{Lepto-Hadronic Emission~Modelling}

While leptonic models represent the most economic approach in terms of free parameters and energy requirements, there are several reasons to consider an additional hadronic component that might contribute to the observed broadband emission. One motivation is the open question on the
origin of ultra-high-energy cosmic rays (UHECRs) and high-energy neutrinos. Observations with current detectors, such as the Pierre Auger Observatory and IceCube, seem to favour an extragalactic origin at least for a fraction of the observed particle fluxes~\cite{Auger2017,IceCube2018}, leaving blazars as one of the few potential emitters, together with other types of AGNs, starburst galaxies or GRBs. A~direct link between astrophysical sources and detected particles is notoriously difficult to establish, but~lepto-hadronic models offer the possibility of connecting a potential site of acceleration of UHECRs with the expected emission of neutrinos and photons up to the VHE~range.  

From a more fundamental point of view, there is strong evidence from observations and modelling that relativistic blazar jets contain an important hadronic component~\cite{Celotti2008}. Depending on the assumed particle acceleration process (see Section~\ref{sec:acceleration}), the~existence of a highly relativistic hadron population may thus seem~natural.

Even within the single-zone framework, different flavours of such models exist. Given the complexity of nuclear interactions and the presumed strong
dominance of hydrogen nuclei, at~least for a large range of energies, most models restrict themselves to the treatment of protons~\cite{Mannheim1993, Aharonian2000, Muecke2001}, although~
models with a more comprehensive treatment of nuclei exist~\cite{Rodrigues2018}. In~the extreme scenario of purely hadronic models~\cite{Mastichiadis2013}, all emission
is traced back to a relativistic population of hadrons. Leptonic synchrotron emission comes only from secondary electrons created in hadronic interactions. Such scenarios require usually extreme jet~powers. 

A more common approach is the
``lepto-hadronic'' framework, where the low-energy emission bump of the blazar SED is ascribed to electron synchrotron emission, as~in leptonic
models, while relativistic hadrons contribute to---and in some cases dominate---the high-energy emission. 

Two distinct regimes in parameter space exist to explain a
hadron dominated high-energy bump (e.g.,~\cite{Mastichiadis2013,Zech2017}), (i)  strong magnetic fields, of~the order of $1$\,G to $100$\,G and large proton Lorentz factors of up to $\gamma_{\rm{p;max}} \approx 10^{10}$ 
lead to a dominant proton-synchrotron emission at high energies, or~(ii) smaller magnetic fields combined with a large particle and/or photon field density result in an emission dominated by synchrotron-pair or IC-pair cascades. For~low magnetic field strengths, mixed solutions can be found where leptonic and hadronic processes contribute both at high energies~\cite{Cerruti2015}. Generally, to~contribute significantly to high energies, the~energy density of relativistic hadrons needs to largely exceed that of relativistic leptons; this could be ascribed to specificities of the acceleration process or to the much faster radiative cooling of~electrons. 

Different hadronic interaction processes contribute to the emission of very energetic photons and to the production of secondary leptons. Such models need to account for synchrotron emission from hadrons and secondary particles (electrons, muons, and~to a lesser extent pions and other mesons), for~pair production in the Bethe-Heitler process and, at~the highest energies, in~photo-pion production processes. A~precise description of the latter requires Monte Carlo simulations based on measured cross-sections or descriptions using parameterisations of such 
simulations. In~interactions of ultra-relativistic protons with target photons or protons, the~emission region is opaque for first generations of secondary particles, including gamma-rays, leading to successive formations of the above mentioned internal particle cascades. If~relativistic
nuclei are included, spallation processes need to be taken into consideration as well. A~more detailed description of all relevant interactions and emission processes can be found, e.g.,~in~\cite{Cerruti2020a} and references~therein.

To reduce the additional number of free parameters in lepto-hadronic models that are required to describe the energy distribution of the hadronic population, often the hypothesis of a co-acceleration of leptons and hadrons, resulting in a similar spectral shape for both populations, is invoked. When introducing a certain number of constraints, a~parameter scan can be undertaken with lepto-hadronic single-zone models to explore the range of accessible solutions~\cite{Cerruti2015, Zech2017}.

\begin{figure}[H]
\includegraphics[width=12 cm]{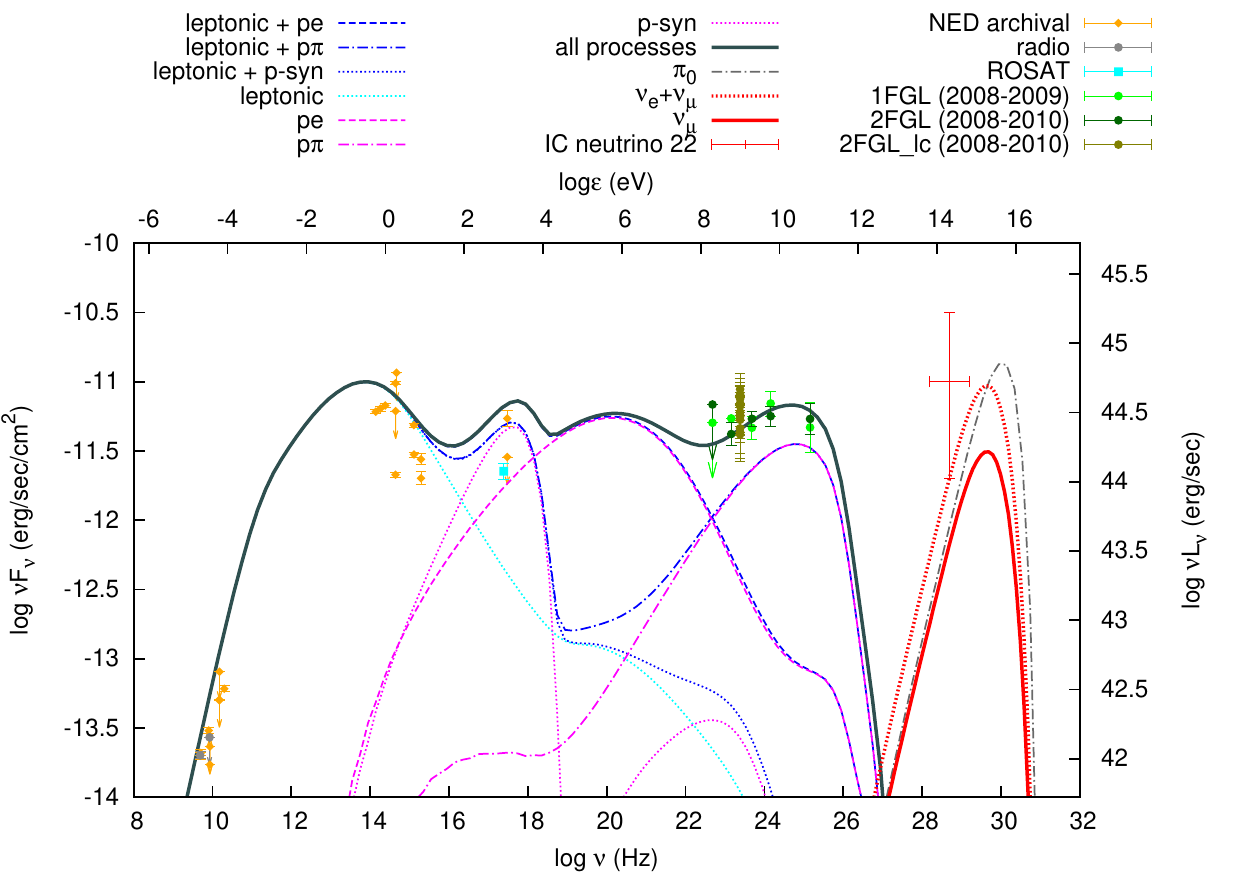}
\caption{Example 
 of a complex lepto-hadronic model, applied to the SED of the BL Lac object 1H\,1914-194, including synchrotron and SSC emission from primary leptons, proton-synchrotron emission, and~radiation from secondary leptons generated in Bethe-Heitler pair production (``pe'') and photo-pion production (``p$\pi$''). In~addition, the~neutrino spectrum emerging from hadronic interactions is shown at the highest energies.
 (taken from~\cite{Petropoulou2015}) \label{fig:leha}.}
\end{figure}

An example for a stationary lepto-hadronic model is shown in Figure~\ref{fig:leha}. In~this particular configuration, proton-synchrotron emission dominates in the X-ray band, while the highest energies are reproduced as the sum of synchrotron radiation from cascades that are triggered by proton-photon interactions (Bethe-Heitler and photo-pion processes). The~significant contribution from the latter leads also to 
the emergence of high-energy neutrinos (see Section~\ref{subsec:neutrinos}). 

Proton-proton interactions are usually negligible in lepto-hadronic blazar models, given the high particle density and the extreme jet powers that a significant emission from such processes would require~\cite{Reynoso2011}. A~possible exception is given by models where the jet interacts with an obstacle, which will be discussed shortly~below.

\subsection{Application of Single-Zone Models to Specific Blazar~Types} 

\subsubsection{Applications to FSRQs and Luminous BL Lac~Objects}

Steady-state emission from luminous flat-spectrum radio quasars (FSRQs) is generally well described by single-zone models when taking into account external photon fields, while both the broad-line region (BLR) and the dusty torus may provide the necessary photon density, detection of VHE gamma rays from such sources puts the emission region
outside or at most at the outer edge of the BLR, otherwise photon-photon absorption would be too significant. Such types of scenarios lead to rather large emitting zones, which have difficulty explaining ultrafast-varying active states (see Section~\ref{subsec:VHE}). As~a result, flare emission from the well-known FSRQ 3C\,279 proves very challenging for single-zone models, be they leptonic or lepto-hadronic (e.g.,~\cite{HESS2019}). 

Intermediate blazar types (LBLs and IBLs, i.e.\ low-frequency peaked and intermediate-frequency peaked BL Lac objects) may also include a significant EIC component (e.g.,~\cite{Boettcher2013}). A~scenario for the LBL AP\,Librae, taking into account information from VLBI data, proposes for instance a contribution from the photon field generated by the extended jet surrounding the high-energy emission region~\cite{Hervet2015}. 
Another model for the same source invokes upscattering of photons from the cosmic microwave background (CMB) by
electrons or pairs in a large-scale jet~\cite{Zacharias2016}.   A~lepto-hadronic scenario is also a possible alternative~\cite{Petropoulou2017}.

However a general study opposing leptonic and lepto-hadronic single-zone interpretations of
steady-state SEDs from FSRQs, IBLs and LBLs~\cite{Zdziarski2015}, concludes  that, unless~we modify our current understanding of accretion and jet launching mechanisms, the~lepto-hadronic interpretation is disfavoured due to the very large required jet powers, largely exceeding the Eddington limit, that are in contradiction with estimates of jet powers by other methods, requiring at the same time an exceedingly low radiative accretion efficiency. 

\subsubsection{Applications to HBLs and Extreme~Blazars}

Blazars of the HBL type are generally very well described with single-zone SSC models, as~long as the data sets are limited to steady states. 
These models require usually
small magnetic fields, in~the range from $0.01$ to $0.1$\,G, far away from equipartition between magnetic and kinetic energy density in the emitting gamma-ray zone. A~representation of the synchrotron bump with a single power-law
electron distribution, including a spectral break with a change of index $\Delta n =1$ as expected from a simple radiative cooling, 
is not possible in most cases. A~broken power-law with a much steeper second spectral part or possibly a logparabolic distribution are required.
This seems to indicate that the basic homogeneous single-zone picture is a simplification,
where a free parametrization of the particle distribution is necessary to model a more complex combination of particle populations and inhomogeneities in the emission region, considering additional effects such as  particle injection, acceleration, and~escape. Lepto-hadronic models always provide an alternative explanation, with~the same caveats as for the luminous blazars. In~the case of HBLs, it is possible to find
solutions with sub-Eddington luminosities, but~the large difference between required jet power and very low disk luminosity still remains puzzling.
As for the more luminous blazars, modelling of flares pushes the single-zone approach, whether leptonic or lepto-hadronic, to~its limit also for HBLs. The~steep flux ratio
between the VHE and X-ray band during the gigantic 2006 flares of PKS\,2155-304
for example, seems to require at minimum a combination of a region dominating the steady-state emission and a more compact region responsible
for the flare emission~\cite{hess2012}. To~reproduce the MWL light curve and
spectral evolution during the 2010 flare of Mrk\,421 also requires contributions from several emission or acceleration regions~\cite{Dmytriiev2021}, to~cite another example among several others.

The validity of the single-zone SSC model is also put into question, when trying to apply it to the most extreme HBLs. Such ``extreme-TeV'' blazars~\cite{Biteau2020}
 exhibit very narrow spectral bumps, peaking in the X-ray and TeV range. To~reproduce their extremely hard spectra requires a very low magnetic field in the $\upmu$\,G range, and~either very hard particle spectra, beyond~what seems acceptable for standard particle acceleration mechanisms, or~the artificial introduction of a very high value of the minimum electron Lorentz factor, on~the order of $10^3$ or $10^4$ (e.g.,~\cite{Katarzynski2006,Costamante2018a}). Acceleration through magnetic reconnection, while able to explain hard particle spectra, requires relatively high magnetization to be efficient (e.g.,~\cite{Giannios2009}), contrary to what is derived from the spectral distributions of these objects. A~physical explanation for such unusual parameters for extreme blazars has been proposed in the form of electron-proton
co-acceleration on relativistic shocks~\cite{Zech2021}. Electrons are seen in PIC simulations to be boosted to higher Lorentz factors, as~interactions with protons lead to an energy transfer to electrons, until~the energies of the two particle species are roughly in equipartition. Low magnetisation is a prerequisite for efficient shock acceleration and arises naturally in this model.   
While technically a lepto-hadronic scenario, the~hadronic emission component is completely negligible, when assuming an equal number density of electrons and protons. To~explain the hardest spectra in this scenario requires re-acceleration on a second shock or on turbulences within the jet~\cite{Tavecchio2022}. Alternative scenarios evoke significant adiabatic losses and specific stochastic acceleration models~\cite{Lefa2011}, different types of lepto-hadronic models~\cite{Cerruti2015}, external cascades in the intergalactic medium~\cite{Tavecchio2014} or EIC on CMB photons along the jet~\cite{Boettcher2008}.


\subsection{Strengths and Weaknesses of Emission~Scenarios}

As presented above, current emission scenarios can usually reproduce the SED observed during quiescent states of blazars. However, two general problems arise in the lepto-hadronic framework. First, if~hadronic processes contribute significantly to the high-energy emission, jet powers are always very large, either dominated by the magnetic energy density in proton-synchrotron dominated scenarios or by the kinetic energy density of the hadrons in scenarios where proton-photon interactions dominate. 
Sub-Eddington solutions can be found for certain sources and scenarios, but~it is generally difficult to reconcile the large required jet power with the accretion power inferred from known processes and from constraints on the luminosity of the accretion disk. Second, most lepto-hadronic models face difficulties in accounting for rapid variability observed at very high energies, due to the much longer acceleration and cooling time scales for hadrons, as~compared to electrons. 
 The characteristic synchrotron cooling time $t_{syn}$ of a particle, for~example, depends on its mass $m$ as:
\begin{equation}
     t_{syn} = \frac{3 m^2 c^3}{4 \sigma_T U_B \beta^2 E}
 \end{equation}
 with $\sigma_T$ the Thomson cross section, $U_B$ the magnetic energy density, $\beta$ and $E$ the
 initial velocity (in units of $c$) and energy of the particle.
 To reach short variability time scales requires scenarios invoking very high magnetic field strengths or possibly jets colliding with matter. 

Leptonic emission models can more easily reproduce the short time scales seen in VHE blazar flares, and~explain the often observed simultaneous variations of the low-energy and high-energy bumps of the SED. However, the~interpretation of very rapid variability down to flux doubling times of only a few minutes, which has been observed in a few exceptional flares~\cite{Gaidos1996, Aharonian2007}, requires extreme parameters or models that differ greatly from the standard view of an emission region embedded inside the jet and with an extension comparable to
the jet radius. Other limitations of current emission models come from certain emission signatures, such as ``orphan'' high-enery flares without counterparts at lower energies (e.g.,~\cite{Krawczynski2004}) or ``extreme'' blazars with very hard spectra in the high-energy band (discussed above) (cf.~\cite{Biteau2020} and references therein), which cannot be described with single-zone leptonic models with standard parameters and generally require lepto-hadronic solutions or more sophisticated multi-zone models. It is also interesting to note that, at~least in the absence of strong external photon fields, both leptonic and lepto-hadronic single-zone models generally provide solutions that are far out of equipartition between the particle energy densities and the magnetic energy density, which can be debatable in particular for the modeling of the stationary~states.


\subsection{Beyond the Single-Zone~Model}

The present lack of strong observational constraints on key parameters such as the magnetic field strength and particle densities of the emission regions or their physical extension leads to a degeneracy of even the simplest single-zone descriptions of the steady emission. 
In the high-energy range, where AGN jets are generally not spatially resolved, 
only high-quality multi-wavelength data of flaring events can break this degeneracy and, in~some cases, reject the simplistic single-zone approach. Conceptually, the~single zone is most commonly
understood as a plasma ``blob'' moving along the jet or as an approximation for a stationary region inside the jet with a plasma flowing through it. 
The emission from particles leaving the single zone is neglected, evoking a strong difference in the physical conditions within and outside of the emission region, such as the magnetic field strength or particle density. 
An observational justification for the single-zone approach can be found in the compact emission regions seen in VLBI images of moving and standing ``radio knots'', which are also seen in the optical and X-ray bands for the most nearby radio galaxies~\cite{Snios2019a, Snios2019b}. The~implicit assumption is that one compact emission region, amplified 
by Doppler boosting in beamed jets, dominates largely the emission from intermediate to very high energies, 
while the large-scale emission from the jet is ignored. Simplistic as they may be, single-zone models have been very successful in describing the steady multi-wavelength emission from most BL Lac type blazars and, with~the addition of external photons fields, for~the more luminous FSRQs. They can also constitute the basic building block for more complex models that may be required to describe emission from flares or extended regions. Describing the physical source with only a few basic parameters---the redshift, radius of the emission region, magnetic field strength and Doppler factor, in~most cases---permits to study the impact of the particle distribution and its evolution with time (e.g.,~\cite{Dimitrakoudis2012, Dmytriiev2021}).

Inhomogeneous jet models propose a more realistic treatment of an extended emission region, which might be identified with the whole of the relativistic jet or a large part of it. The~more complete description of the source geometry and its physical parameters is often done at the expense of a less detailed characterisation of the particle distribution. Models beyond the single zone are clearly needed when emission is thought to be extended along the jet and when one wants to describe the different acceleration and cooling time scales of electrons injected into the jet at different energies and propagating along it, leading to non-trivial multi-frequency variability and polarization patterns (e.g.,~\cite{Boutelier2008, Potter2013, Marscher2014, Lucchini2019}).

The simulation of polarization in the radio and optical band in multi-zone or MHD models clearly provides a very useful additional observable that can be verified against observations to probe the structure and degree of order of the magnetic field in different regions of the jet (for a recent comprehensive review see \cite{Aller2021}). Temporal changes in the direction of the optical polarization angle have also been identified with gamma-ray flares, providing clues to the structure and location of the emission regions of flares, thus further constraining emission scenarios (e.g.,~\cite{Fermi2010, Blinov2018}).
Apart from probing the magnetic field structure, future polarization studies in the X-ray and gamma-ray band, may also show some promise in distinguishing between leptonic and (lepto-)hadronic emission models~\cite{Zhang2013}. This is due to the fact that the high-energy emission in leptonic models is dominated by Inverse Compton emission, leading to a lower degree of polarization than the synchrotron emission expected from proton primaries or from secondary cascade particles in (lepto-)hadronic~models.

It should be noted that the treatment of the internal light-travel time effects becomes important 
when dealing with an extended emission region~\cite{Chiaberge1999}.
This more complex multi-zone approach, when extended to lepto-hadronic modelling, requires in addition the follow-up of several particle species over a large range of simulated cells (e.g.,~\cite{Zacharias2022}).

A particular example of inhomogeneous scenarios is given by the spine-in-sheath model, where the jet is modelled as a fast moving spine, often consisting of an electron-positron pair plasma, within~an often baryonic, less relativistic, sheath layer (e.g.,~\cite{Sol1989, Tavecchio2008, Sikora2016a, Vuillaume2018}). This scenario has been initially proposed to reconcile the high bulk Lorentz factors of a few 10~s inferred from radiative modelling of gamma-loud blazars, with~the significantly lower values derived from direct observations of the movement of VLBI radio knots. It can also reconcile very fast variability from a thin spine with a more slowly varying emission from the sheath. Low-energy emission from the surrounding sheath creates also an additional photon field that can add to the EIC component produced by electrons inside the spine. 
Even if jets from AGNs seem to carry an important fraction of hadrons, the~exact lepton to hadron ratio is an open question. To~reconcile jet power estimates from radiative modelling with much lower observational estimates from studies of X-ray cavities and the energy input into radio lobes, a~contribution from electron-positron pairs seems to be required~\cite{Sikora2016b}. This is another indication for a structured jet that contains a pair plasma in addition to a baryonic~component.

Multi-zone models can also make it possible to consider a separate (purely leptonic) origin for rapid flares, while still allowing for a lepto-hadronic solution for the steady emission. However, the~case of very rapid variability during exceptional flares, with~flux doubling time scales down to a few minutes, most strikingly observed in the HBL PKS\,2155-304~\cite{Aharonian2007} in 2006, still remains a general problem for current models.
Among several propositions to explain these observations are ``minijet''~\cite{Giannios2009,Nalewajko2011} and ``needle-in-jet''~\cite{Ghisellini2008} scenarios, where the existence of multiple, very compact emission regions inside the jet, induced by magneto-centrifugal acceleration of electron beams or magnetic reconnection in a Poynting flux dominated jet, try to explain the inferred small size and very large Doppler factor of the emission region behind the flares.
One way to incorporate very rapid variability in lepto-hadronic models of extended jets is by assuming collisions with clouds, possibly from the broad-line region, or~stellar envelopes (e.g.,~\cite{Barkov2012,Bosch-Ramon2012}).

\begin{figure}[H]
\centering
\includegraphics[width=0.57 \columnwidth]{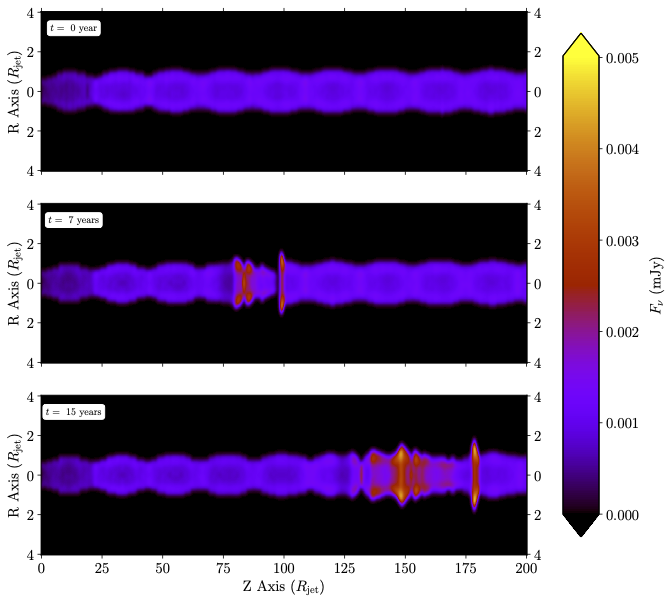}
\includegraphics[width=0.42 \columnwidth]{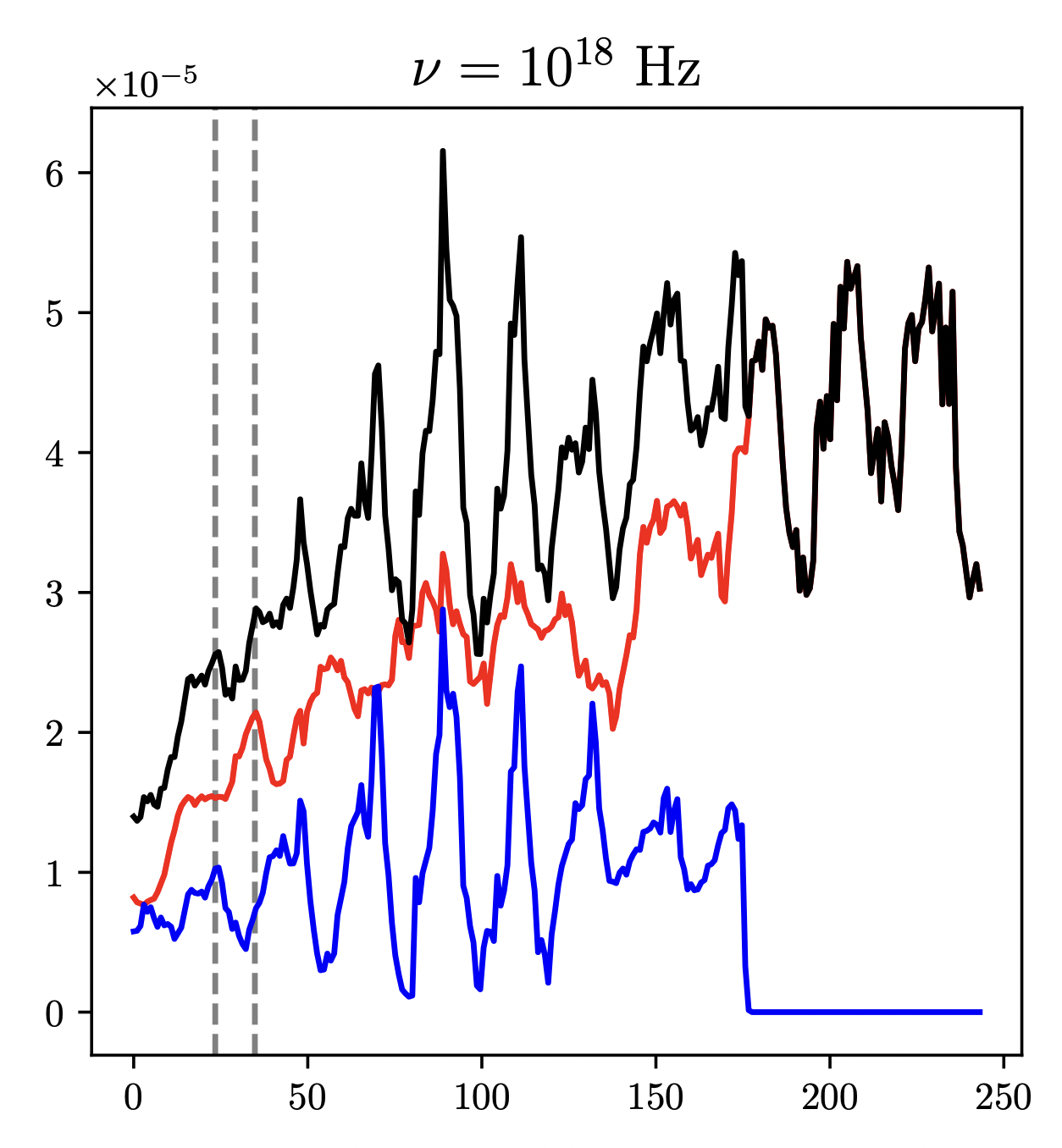}
\caption{({\bf Left}): Synchrotron emission map at $10^{10}$\,Hz of an MHD simulation showing
the impact of a fast ejecta propagating through a series of recollimation shocks in a relativistic jet, seen under $90^{\circ}$. The~x- and y-axes are given in units of the jet-radius $R_{jet}$. ({\bf Right}): Light curve of the emission in the X-ray range showing the contribution from the ejecta (blue) and the remaining jet (red). The~units on the X-axis are years in the observer frame (taken from~\cite{Fichet2022}). Credit: \cite{Fichet2022}, A\&A, reproduced with permission @ESO.
\label{fig:mhd}}
\end{figure}

Taking the description of the jet physics one step further, several models include emission processes in magneto-hydrodynamic (MHD) and particle-in-cell (PIC) simulations of relativistic jets, while such an approach---computationally expensive of course---leads to a large extension of the parameter space, it carries the promise of mutually constraining the macrophysics of the jet evolution inside the ambient medium and the microphysics of the naturally arising acceleration and emission regions inside the jet. Such models try to directly link the observed emission with the assumed acceleration processes inside the jet, such as acceleration on moving and recollimation shocks---often limited to a link with synchrotron emission in the radio domain---e.g.,~\cite{Gomez1995,Agudo2001,Mimica2009,Porth2011,Fromm2016,Fichet2021}, shear acceleration (cf.~\cite{Rieger2019} and references therein), or~magnetic reconnection (e.g.,~\cite{Petropoulou2019}).
Figure~\ref{fig:mhd} illustrates for example a synthetic radio-synchrotron map of a simulated jet, where the appearance of MWL flares is attributed to shock acceleration of electrons or pairs in collisions of moving and stationary shocks, as~observed on kpc-scale in the jet of 3C264~\cite{Meyer2015}. 
Such a combination of different approaches seems best suited to exploit at the same time the kinematic and dynamic information on the jet one gains from time-resolved VLBI observations and the information on acceleration and emission processes that are most strongly constrained by the highly variable gamma-ray~emission. 

\section{Brief Overview of Particle Acceleration~Mechanisms}
\label{sec:acceleration}

All current emission models require the presence of a population of charged particles accelerated up to very high energies, which are the true origin of the VHE radiation. However the dominant acceleration mechanisms at work to generate such populations in VHE blazars are not yet well identified. Several possibilities are open as illustrated in Section~\ref{sec:models}. They have been extensively studied in the literature and are still the subject of in-depth and constantly improving analyzes due to the exponential growth in the power of supercomputers. We only recall here the main ones, supposed to play a major role in AGN jets.

\subsection{Fermi Processes and Shear~Acceleration}

Turbulence and shocks are naturally expected to develop in the collisionless plasma of astrophysical jets, providing a wide range of possibilities for activating Fermi-type acceleration mechanisms where particles can reach very high energies through their successive interactions with moving magnetized scattering~centers. 

The growth of Kelvin-Helmholtz instabilities for instance at the interface of the jet with its surrounding ambient medium, or~at internal interfaces between various jet layers or knots with different bulk velocities in stratified jets,  can generate turbulence and induce type II Fermi acceleration. Relativistic particles gain energy with an average gain of $<\Delta E> \propto (u_c/c)^2 E$ per interaction, where $E$ is the energy of the particle and $u_c$ the velocity of the scattering center. The~characteristic acceleration time $\tau_{acc}$ is independent of $E$ and a population of particles with a differential distribution $dN(E) \propto E^{-p} dE$ can be generated with a power law index $p = 1 + (\tau_{acc}/\tau_{esc})$ if $\tau_{esc}$, the~characteristic escape time from the acceleration zone, is also energy-independent. Such type of stochastic acceleration has been shown to generate very high energy particles, both in the quasilinear theory and in the regime of strong nonlinear turbulence~\citep{Schlickeiser2002,Stawarz2008,Zhdankin2017,Zhdankin2018,Comisso2018,Nattila2021,Bresci2022}, and~is expected to play a role in quiescent and flaring blazars~\citep{tramacere2011,Dmytriiev2021,Asano2015}. 

The presence of shearing flows inside the jet and at its periphery can even enhance the efficiency of such processes by the so-called ``shear acceleration''. It was analyzed that instabilities and turbulence induced by shearing in an outer layer of thickness of the order of $0.1 R_{jet}$ around a jet of radius $R_{jet}$ can accelerate electrons up to PeV energies and protons up to EeV energies, without~destroying the jet by instabilities~\citep{Rieger2021}. The~average gain of energy per interaction by a particle can be expressed as $<\Delta E> \propto (u/c)^2 E \propto (l/c)^2 (\partial {v_z} /\partial x)^2  E$ for a flow with speed $v_z$ along the $z$-axis sheared in the transverse direction $x$, where $u = l(\partial {v_z} /\partial x)$ is the effective speed change of the scattering centers as seen by a particle crossing the flow along the $x$-axis on one mean free path $l$. The~acceleration time being inversely proportional to the mean free path, $\tau_{acc} \propto c/l (\partial {v_z} /\partial x)^2$, shearing acceleration appears faster for particles already accelerated at high energies, and~more efficient for protons with large $l$ than for electrons. The~distribution of accelerated particles tends to a power law with index depending on the properties of the underlying turbulence~\citep{Rieger2019}. 

A wide variety of shocks are also expected to develop in supersonic AGN jets, such as reconfinement shocks, stationary or moving internal shocks along the jet, or~large scale bow shocks related to hot spots at the extremity of Fanaroff-Riley type II radiosources. 
All these shocks are privileged sites to launch the activation of type I Fermi acceleration mechanisms~\citep{Spitkovsky2008}. Fast particles can be partially trapped in the vicinity of the shock by their scattering on the magnetic turbulence that they themselves 
amplify, and~cross non-relativistic shocks several times with an important average gain of energy for each round trip $<\Delta E> \propto (\Delta u/c) E \propto [(r-1)/r] (u_{shock}/c)$, where $\Delta u$ is the relative velocity between the upstream and downstream medium, $u_{shock}$ the shock speed in the frame of the upstream medium, and~$r$ the shock compression ratio, of~the order of 4 for strong very supersonic shocks. Here the ratio $\tau_{acc}/\tau_{esc}$ is energy-independent because the same scattering mechanisms are responsible for both particle acceleration and escape, and~the differential distribution of particles is then a power law with index $p = (r+2)/(r-1)$ which is of the order of 2 for strong shocks, and~presents a rather universal character often observed in the cosmos and obtained in numerical~simulations.

The case of ultra-relativistic shocks which can develop in AGN jets with a bulk Lorentz factor $\Gamma_{shock} >> 1$ is quite special because only energetic particles with individual Lorentz factor $\gamma >> \Gamma_{shock}$ can successfully cross the shock from downstream to upstream. Moreover in the case of magnetized shocks such crossing from downstream to upstream is possible only for parallel or quasi-parallel shocks, with~a small angle between the magnetic field and the flow direction. Type I Fermi processes are then much less efficient, especially in perpendicular and quasi-perpendicular shocks, and~typical spectral indices are then softer, of~the order of $2.2$ to $2.3$ 
 as estimated by semi-analytical methods, and~Monte Carlo (MC) or Particle in Cell (PIC) simulations~\citep{Nishikawa2005,Spitkovsky2008,Sironi2011,Summerlin2012,Ellison2013,Guo2014,Sironi2015,Groselj2022}.  However, shock surfing and shock drift acceleration can still develop in quasi-perpendicular shocks because of the non-zero electric field in $V \times  B/c$ induced by particle motion towards the magnetized shock. Such mechanisms sometimes described as ``fast Fermi processes'' allow particles to increase their energy by an order of magnitude in a single encounter with the shock. For~weak levels of turbulence, shock drift acceleration can become dominant in oblique shocks and generate flat-spectrum population of energetic particles up to the highest energies as obtained in MC simulations. Acceleration in relativistic shocks can then produce populations of VHE particles with a large range of power-law indices, from~hard index $p \simeq 1$ to very steep ones, depending critically on the properties of turbulence and magnetic field, on~the shock speed, and~shock obliquity for mildly relativistic shocks ~\citep{Summerlin2012}. The~coupling of such MC simulations with radiation transfer models made it possible to reproduce quite well the typical flaring behavior of the two archetypal blazars 3C 279 and Mrk 501~\citep{Bottcher2019}. One difficulty with the MC approach resides in the powerful test particle approximation and the fact that feedback mechanisms of the accelerated particles on the ambient plasma, shock and turbulence are mostly neglected. This can be questionable for substantial populations of non-thermal particles. Conversely first principles PIC simulations provide a self-consistent description of the microphysics which reduces the efficiency of the Fermi acceleration in relativistic shocks likely due to the growth of instabilities and  micro-turbulence which reduces the trapping of particles near the shock. Multiple shock re-acceleration scenarios could then provide a way out to explain the hardest gamma-ray spectra observed in extreme TeV blazars~\citep{Zech2021}. However, due to its high computational cost, the~PIC approach also has its own limitations. In~particular it does not yet describe the long term evolution of the shock and large scale phenomena, and~has difficulties to reach the high individual particle Lorentz factors needed to reproduce VHE spectra. 

Interestingly it has been shown that the different aspects of Fermi-type acceleration processes can be described within a same formalism applicable to sub- or ultra-relativistic cases, turbulent, shear and shock acceleration, in~flat or nonflat spacetime~\cite{Lemoine2019}. Such types of approach could make it possible to better identify the domains of predominance of the various Fermi acceleration regimes, including also the immediate vicinity of massive black~holes.

\subsection{Magnetic~Reconnection}

Breaking and reconnecting magnetic field lines can release substantial amount of magnetic energy transmitted to the plasma in various forms, including significant particle acceleration. The~time evolution of the magnetic field written in the context of MHD illustrates this type of phenomena, $\partial B/ \partial t = \bigtriangledown \times (u \times B) - \bigtriangledown \times \eta (\bigtriangledown \times B)$, where $\eta = c^2 / 4  \pi \sigma$ is the plasma magnetic diffusivity, with~$\sigma$ its conductivity. In~the highly conducting astrophysical plasmas, the~first term on the right hand side generally dominates the magnetic field behavior, the~magnetic field is frozen in the plasma and no reconnection occurs. However the diffusive term can become dominant in various situations, for~instance when the first term dwindles to zero like around stagnation points, lines or surfaces with $u \simeq 0$. The~magnetic field then diffuses in a time $\tau_{diff} \simeq {L_B}^2/\eta$, $L_B$ being the spatial scale of magnetic field variations, possibly inducing strong currents and current sheets with non-zero electric field, resulting in plasma heating and particle acceleration. Growing instabilities and turbulences induced by the currents can then lead to the development of an anomalous resistivity which further amplifies the dissipation of the magnetic field. The~study of faster and even explosive magnetic reconnection events, which can occur in hot and diffuse plasmas, requires describing the microphysics where ions and electrons decouple and new terms appear especially in the Ohm's law~\citep{Treumann2015}. Numerical  simulations based on fundamental kinetic Vlasov-Maxwell equations show that relativistic magnetic reconnection can then accelerate particles up to very high energies with a power law distribution with index $p \simeq 1.5$ to $1$ for high magnetisation ~\citep{Marcowith2020,Nishikawa2021}. Despite the fact that the various aspects of magnetic reconnection are not yet fully understood, there have been great advances in the field thanks to numerous theoretical works and a large number of studies presenting simulations 2D and 3D PIC simulations of relativistic magnetic reconnection~\citep{Zenitani2008,Sironi2014,Sironi2016,Werner2017,Christie2020,Nishikawa2020,Zhang2021a,Sironi2021,Meli2022} (see~\citep{Marcowith2020,Nishikawa2021} for  reviews). Such  phenomena should naturally occur in AGN jets where various instabilities can grow, with~complex and turbulent velocity fields, eddies and sheared flows, and~are also observed in simulations of black hole magnetospheres~\citep{Bransgrove2021,Crinquand2022,Nathanail2022}. They attract more and more attention to explain AGN flares and variabilities~\citep{Giannios2009,Nalewajko2011,Morris2019,Asano2020,Sobacchi2021,Medina2021}.

\subsection{Pulsar-like Acceleration around Black~Hole}

Like pulsar magnetospheres~\citep{Goldreich1969}, it seems that the magnetosphere of rotating black holes embedded in a magnetic field can very efficiently accelerate particles, either by centrifugal forces present throughout the magnetosphere~\citep{Gangadhara1997,Osmanov2007}, or~by strong electric fields which grow in gaps, voids of plasma which can develop around critical surfaces in the vicinity of the black hole~\citep{Beskin1992,Hirotani1998}. Particles are expected to be present in the magnetospheres, possibly generated by pair creation or injected from the turbulent accretion flow and disk. For~weak enough ambient radiation field and typical magnetic field of a few tens of Gauss, they can then reach high energies at the outer light surface, estimated to possibly reach 10 to 100 TeV for electrons and positrons, and~up to $10^{18}$ to $10^{20}$ eV for protons which do not suffer high synchrotron losses~\citep{Neronov2007,Rieger2008,Istomin2009,Osmanov2010,Broderick2015,Katsoulakos2018,Parfrey2019,Crinquand2020,Katsoulakos2020,Istomin2021}. Such effects have been proposed especially to explain ultra-fast varying AGN flares (see Section~\ref{sec:zone}) and should at least contribute to inject a population of energetic particles at the base of jets and winds from the central AGN engine (e.g.,~\cite{MAGIC2014,Broderick2015,Katsoulakos2020}). 

A combination of the different acceleration processes mentioned above is possible and even likely, pending a full understanding of the main accelerator. Different acceleration zones and mechanisms could contribute depending on the local plasma properties such as velocity fields, turbulence, magnetization. Magnetic reconnection requires high magnetization and could significantly contribute for instance at the jet base which is magnetically dominated for jets extracted by the Blandford-Znajek process, while efficient shock acceleration which requires lower magnetization could dominate further in the jet, when enough magnetic field has been dissipated. Indeed it is already often considered that pre-acceleration by magnetic reconnection is at work inside jets and provides particles energetic enough to efficiently interact with plasma waves and structures in Fermi-type acceleration zones. Finding the location of the emitting zone should help identify the dominant acceleration mechanisms and the relative importance of leptonic versus hadronic~processes.

\section{Locating the VHE Emitting Zone: A Critical Missing~Link}
\label{sec:zone}
\unskip

\subsection{Black Hole~Magnetospheres}
Very fast variability detected in some blazars as PKS2155-304 and Mrk501 and in the radiogalaxies M87 and IC310 at TeV energies suggests from the causality argument that the size of the VHE zone should be comparable to or even less than the size of their black hole horizon, even when assuming rather high jet bulk Lorentz factors. Compact black hole magnetospheres where particle acceleration can be very quick and efficient then appear as a natural site for the origin of highly variable VHE flares, depending on the relative importance of the $\gamma$-$\gamma$ absorption phenomena. In~the presence of bright accretion disks usually found in FRSQs, such absorption is expected to drastically limit the propagation of VHE photons. However absorption effects can be overcome in AGNs like BL Lac objects or radiogalaxies with sub-luminous disks in the ADAF regime and sufficiently small magnetic~fields.

\begin{figure}[H]
\includegraphics[width=10.5 cm]{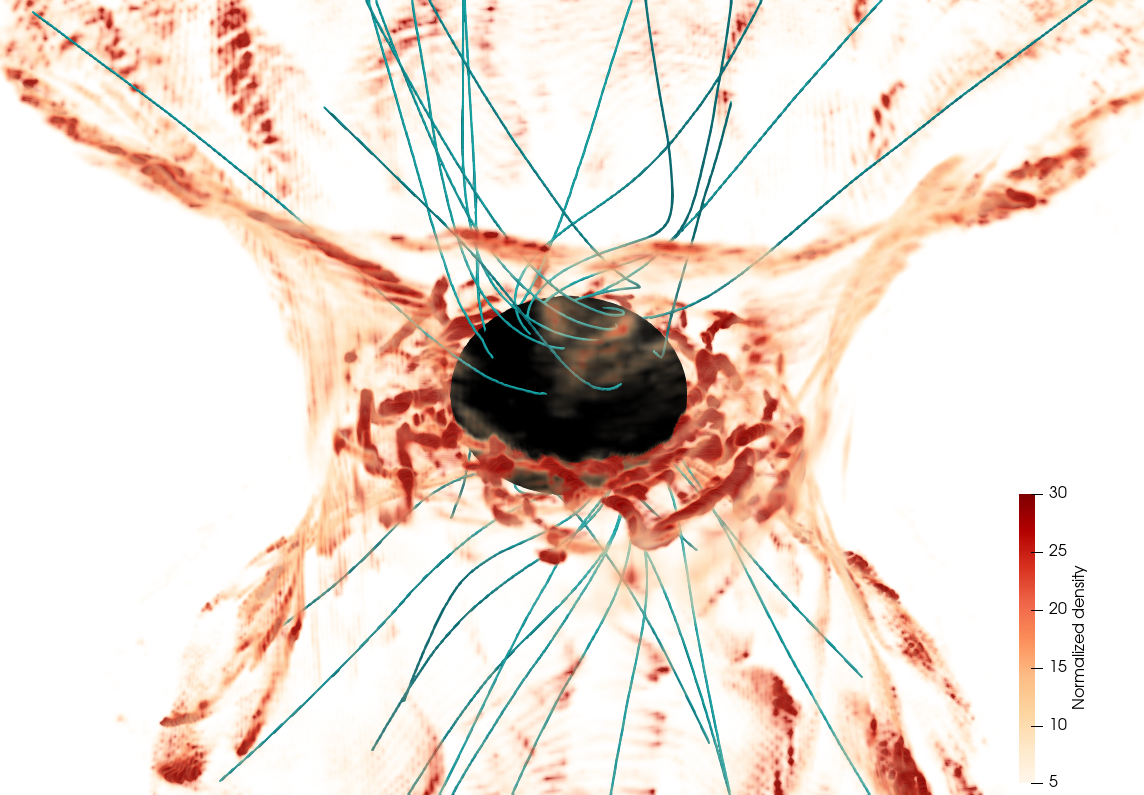}
\caption{3D general relativivistic particle-in-cell simulations (GRPIC) of a rotating black hole magnetosphere. Snapshot of the normalized total plasma density (in brown) with some magnetic field lines (in blue) at $t = 40~{r_g}/c$, showing a persistent fragmented equatorial current sheet where particles are accelerated by magnetic reconnection, very close to the black hole horizon (in black). Taken from~\cite{Crinquand2022}. \label{fig:BHmagneto}}
\end{figure}

Advanced instruments such as the Event Horizon Telescope and GRAVITY are now able to detect very fine structures in the core of nearby AGNs, which recently motivated massive numerical simulations of black hole magnetospheres, using PIC codes in general relativity as illustrated in Figure~\ref{fig:BHmagneto}. In~a quasi force-free approximation, they show the launching of relativistic jets by the Blandford-Znajek process, as~well as the formation of a significant current sheet with magnetic reconnection events in the equatorial plane of a black hole in extreme rotation~\citep{Parfrey2019}. Considering a plasma of electrons, positrons and photons, and~taking into account radiation transfer with inverse-Compton scattering and pair creation, GRPIC simulations can explore the formation of gaps where fast particle acceleration occurs due to the unscreened electric field, together with the generation of high energy radiation in the magnetosphere~\citep{Crinquand2020,Crinquand2021}. An~intermittent gap develops at the base of the jet near the inner light surface for a certain range of optical depth, and~ejects electron-positron plasmoids which induce fast-varying gamma-ray emission. Although~the amplitude of these self-consistently generated bursts appears at first sight to be slightly lower than the amplitudes of the gamma-ray flares currently observed from TeV-emitting AGNs, such a phenomenon deserves attention because it could allow us to probe at VHE the immediate vicinity of the supermassive black hole's horizons. Sudden changes in the magnetosphere parameters, such as the local accretion rate, could possibly enhance the observed variabilities~\citep{Crinquand2021}. Strong and rapid TeV flares dominated by curvature emission of particles accelerated in a spark gap developing at a few gravitational radius $r_g$ from the horizon could be launched for instance by abrupt changes of the disk emission or of magnetospheric currents~\citep{Kisaka2022}. 

\subsection{Nuclear VLBI~Jets}
\label{subsec:nuclearjets}

The base and the inner parts of the jets observed as radio core and knots by VLBI on pc and sub-pc scales are often considered as the most natural place for the VHE emitting zone since it ensures that observed fluxes and variabilities can be directly enhanced by Doppler boosting if the VHE zone is embedded in the relativistic jet bulk outflow. A~wide variety of emission models make this assumption over years, either explicit or implied. From~our current views on hadronic and leptonic models, one can think that leptonic scenarios with magnetic field in the range of $0.01$ to $0.1$~G could apply to VHE emitting zones along the VLBI jets while hadronic scenarios which require much higher magnetic field could concern the base and very inner parts of the jets, possibly inside still unresolved radio~cores.

\begin{figure}[H]
{\captionsetup{position=bottom,justification=centering}
\subfloat[]{
        \includegraphics[clip,width=13.5 cm]{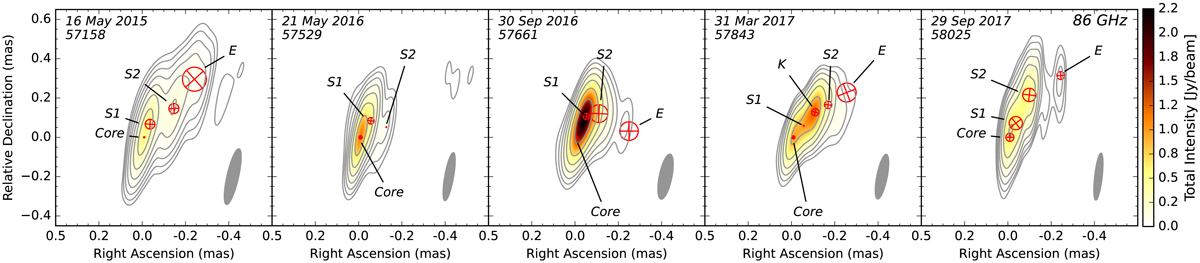}
}
 
\subfloat[]{
        \includegraphics[clip,width=9.5 cm]{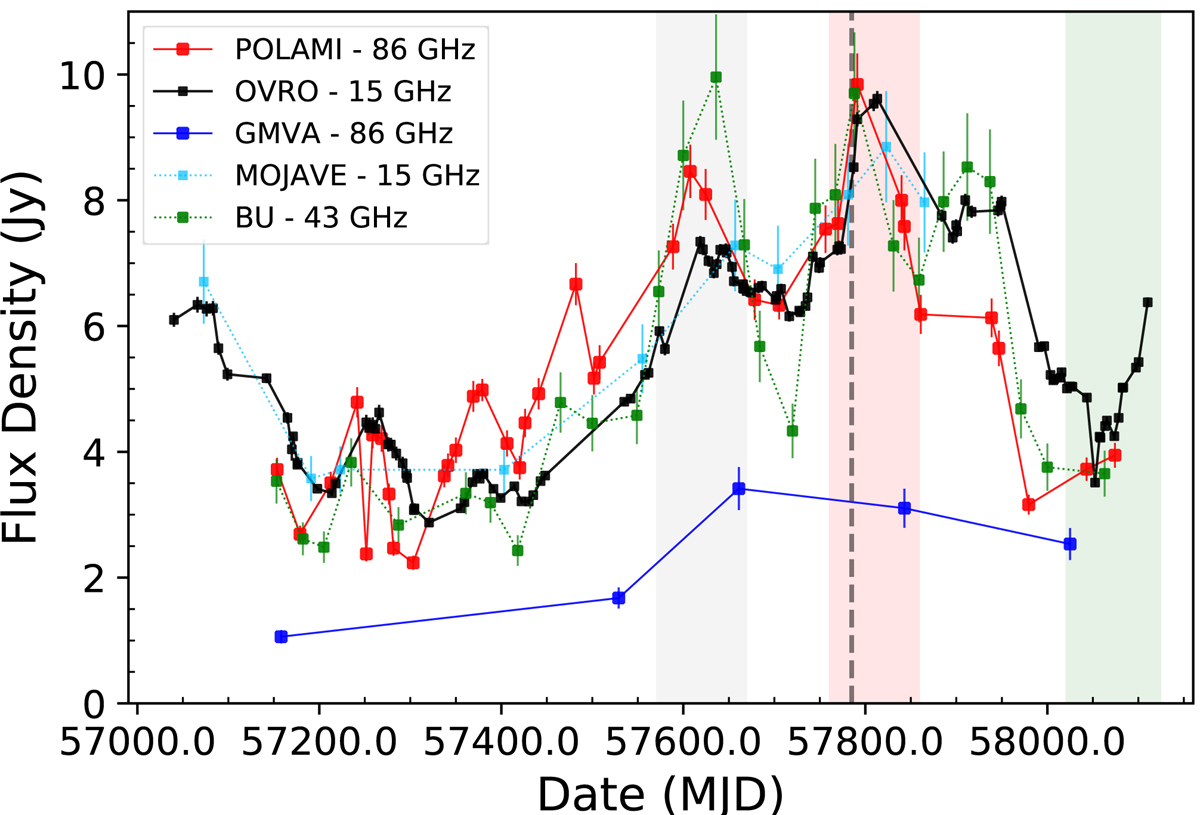}
}}

\caption{Time 
 evolution of the blazar OJ 287 on parsec-scale, showing the appearance of a new VLBI feature (knot K) in the jet following the first ever VHE flare detected from this source during the period  1--4 February 2017, and~the occurrence of a high radio emission state around the TeV activity. ({\bf a}) Total intensity VLBI images at five epochs from 86 GHz GMVA observation. Red circles represent the model fit components. The~new feature K appeared in March 2017, at~~ 0.2 mas from the radio core, between~the two quasi-stationary components S1 and S2. ({\bf b}) Multiwavelength radio light curves from 2015 to 2017. The~vertical dashed line shows the time of the VHE flaring state and shaded areas the periods of enhanced activity. The~TeV detection was also coincident with increased activity in the X-ray and gamma-ray ranges. Credit: \cite{Lico2022}, A\&A, reproduced with permission @ESO.
\label{fig:OJ287}}
\end{figure}

Indeed, there is growing evidence for correlation between some HE and VHE flares and various phenomena detected in radio VLBI such as variations in the core, emergence of new knots at the jet base, or~appearance of stationary features~\citep{Akiyama2015,Jorstad2016,Rani2018,Larionov2020,Kim2020,Lico2022}, which proves that the emission of at least a part of the flares is related to the nuclear jet structure (see Figure~\ref{fig:OJ287}). However the causal link has not yet been fully established, and~the statistical significance for the whole blazar population is still difficult to~assess.

Search for specific variability patterns in HE and VHE light curves is a promising way to further explore possible underlying signatures of the VLBI structure, especially in HBLs which often show a series of stationary VLBI knots possibly interpreted as shock diamonds where particle acceleration and enhanced radiation can occur. Some clues suggesting that a perturbation moving at constant speed along the jet induces a series of flares when crossing the successive shocks have been found for instance in the case of Mrk 421, where the main emission zone would be at about 10 pc from the core, with~secondary zones further down in the jet~\citep{Hervet2019}. This could be confirmed if a unique variability pattern was detected in X-rays or gamma-rays after each strong flare. Such a temporal coincidence between the occurrence of a bright flare and the interaction of a fast-moving VLBI knot with a stationary jet feature has been also recently observed in the FSRQ PKS1510-089~\citep{HESS2021}. If~that correlation is true, the~VHE fast-varying flare comes from a zone located at about 50 pc from the core, outside the BLR, which is consistent with the lack of clear BLR absorption features in the spectrum but raises the question of the origin of the rapid variability. 

\subsection{Large Scale~Jets}

Recent gamma-ray observations of the radiogalaxy Cen A that resolved its radio and X-ray jet at VHE definitely prove that large scale AGN jets can radiate in the TeV domain and suggest that in situ acceleration of ultra-relativistic electrons can be at work over several kpc along AGN jets~\citep{hess2020}. Such  extended VHE emission is often ascribed to comptonization process of various types of soft photon background~\citep{Stawarz2003,Bednarek2019}. In~some cases the geometry of the inverse-Compton scattering process can even lead to large angle gamma-ray emission, for~instance if an isotropic distribution of relativistic electrons from the decelerated kpc-scale jet interact with low energy photons from the inner relativistic jet. Beyond~blazars, unbeamed large scale jets could therefore also produce significant VHE fluxes, reachable by present IACTs for the brightest ones, or~by~CTA. 

Such large scale emitting jets could in principle explain or contribute to the stationary VHE fluxes detected from blazars the gamma-ray emission of which is not angularly resolved, but~they are poorly adapted to explain any fast variability. This shows in filigram that at least two types of VHE emission zones are present in AGN, although~blazars and generally all ultrafast AGN outbursts are likely dominated by compact Doppler boosted emitting zones.   

\subsection{Constraints from VHE~Observations}  
\label{subsec:VHE}

The $\gamma$-$\gamma$ absorption effect can impose severe constraints on the location of the VHE emitting zone in the presence of strong ambient photons with energy $\epsilon_{bkgd} \gtrsim {m_e}^2 c^4/ {\epsilon_{\gamma}}$ since gamma-rays with energy $\epsilon_{\gamma}$ can not escape when the optical depth due to pair creation on ambient photons becomes too important. Although~possibly negligible for BL Lac objects, this effect should drastically affect FSRQs with bright disks, dusty tori and BLR, and~impose some minimal distance of the VHE emitting zone from the central engine and the BLR. This suggests that the emitting gamma-ray zone is located beyond the BLR in FRSQs and appears coherent with studies searching for absorption effects in the gamma-ray spectra of blazars, which detect very few features of absorption by the BLR~\citep{Costamante2018b} but find some signatures of absorption by the more extended narrow line region (NLR) in the sub-TeV range~\citep{Foffano2022}. Conversely, it is puzzling to note that correlation was found between the variations of BLR emission lines and gamma-ray flux in some AGN such as PKS1510-089~\citep{Isler2015}, which suggests on the contrary that part of their HE emission comes from  Inverse-Compton scattering with ambient photons from the~BLR. 

Moreover, as~mentionned in Section~\ref{sec:models}, ultrafast variability appears rather frequent in blazars and radiogalaxies and still remains hard to grasp despite the development of several scenarios (see for instance~\citep{Aharonian2017}). A~strong tension therefore remains on the determination of the location of the VHE emitting zone especially for fast-varying FSRQ such as 3C 279 or PKS 1222+216 for which the causality argument provides a severe upper limit on the size of the emitting plasma while absorption effect constrains the emission to come from rather large spatial scales, far from the central engine. Two main alternatives can be considered for such sources. Specific geometry of the BLR could allow HE photons to escape from the compact core and jet base if the jet is shielded from external photons by a plasma sheath~\citep{Meyer2019}. Another possibility is to assume that the fast-varying flares come from further down the outflow,  from~active small sub-structures in a more extended jet~\cite{Ghisellini2008, Giannios2009, Narayan2012}. More exotic alternatives such as the existence of axion-like particles could also solve the puzzle by allowing VHE gamma-ray to escape from the black hole vicinity below the BLR due to a significant reduction in $\gamma$-$\gamma$ absorption.

The new generation of VHE gamma-ray instruments such as the future imaging atmospheric Cherenkov Telescope Array (CTA), together with the VHE wide field detectors HAWC, LHAASO and SWGO, will detect much larger samples of BL Lacs and FRSQs, with~better temporal and spectral coverage and resolution. They will search for specific spectral features and especially for additional components in the multi-TeV range which can be expected in the presence of hadronic cascades. They should also make it possible to gather many complete and detailed datasets on blazar gamma-ray flares, which are currently very few. Search for periodicities and quasi-periodicities, analysis of the power spectral density of VHE lightcurves, of~possible lognormality and of the nature of the noise for different activity levels, studies of time delays, spectral features and  hysteresis for various sources,  should better constrain the location, geometry and kinematics of the emitting zone(s), as~well as the main leptonic or hadronic scenarios. For~example, the~detection or not of time delays in the arrival times of VHE photons as a function of their energy should strongly constrain the origin of flares in single-zone models, often interpreted as due to competing effects between the particle injection and acceleration and the radiative cooling, which generally induce intrinsic spectral lags, although~there is still no firm detection confirmed by current instruments~\citep{Perennes2020,Levy2022}. The~persistent non-detection of such time delays could be an argument in favor of flare scenarios based on a simple change in the bulk Doppler factor of the emitting zone, which have been proposed in some sources for instance in the case of helical jets~\citep{Roy2022} and in the context of strong recollimation shocks~\citep{Hervet2022}. The~exploration of very fast sub-minute time scales accessible thanks to the expected jump in sensitivity promises to be particularly interesting and should drastically constrain the properties of the emitting zones and emission processes~\citep{Zech2019}. Deep VHE analysis and long-term monitoring of archetypal sources of HBL, IBL, LBL and FSRQ populations,  of~the extreme UHBL blazars which challenge the current scenarios, and~of some peculiar sources such as the binary black hole system OJ 287~\citep{Lico2022,OBrien2017} or the gravitationally lensed QSO B0218+357~\citep{Acciari2022} should further characterize links between the VHE properties of the AGN and the global characteristics and fine structure of their central core and jet.

\section{Multi-Wavelength and Multi-Messenger~Astrophysics}
\label{sec:mwl} 

Beyond analysis of unidentified VHE sources off the galactic plane, which may reveal as yet unknown types of extragalactic cosmic sources, the~MWL and MM approach will be essential to characterize the general  properties of high energy AGN, to~generate a still-missing  sample of tracked flares with good temporal and frequency coverage, and~to  firmly break the degeneracy between the different versions of plausible models of gamma-ray blazars, and~in particular between the leptonic and hadronic emission scenarios. The~current development of large simultaneous campaigns and monitoring, alert networks, ToO programs and data-driven detection appears mandatory to optimize the scientific return of many experiments and large infrastructures catching blazar data. We just mention a few examples~below. 

\subsection{The Infrared, Optical, X-ray and Low-Energy Gamma-Ray~Domains} 

Non-thermal emission is usually expected to dominate in blazars, especially in BL Lacs, and~in the gamma-ray domain. However both non-thermal and thermal emission can be detected in the lower energy ranges. Although~not always easy to disentangle, the~two of them  are important to monitor in order to better constrain and describe, respectively, (i) the populations of HE and VHE particles, (ii) the AGN environment with its ambient photon fields (disk, corona, torus, BLR, NLR, clouds, gaseous and stellar components, host galaxy), and~to deduce the causal links between the different AGN~components.

Several outstanding projects and infrastructures will explore the sky from IR to X-rays in the coming decades, such as the JWST, Vera Rubin Observatory (LSST), ELT, Athena and many others. They will offer plenty of opportunities to conduct deep and wide surveys as well as short and long term MWL monitoring of AGN, making possible the statistical analysis of correlations and time lags between different frequencies. This should characterize the populations of emitting particles and identify the sequence of processes, shedding critical new light on the long-standing question of the origin of blazar variability.
High-quality spectra together with polarimetric monitoring  will be crucial to explore the nature of the optical and X-ray emission, probe the magnetic fields, and~disentangle its various components for the different types of sources~\citep{Aller2021}. Spectra will also be requisite to complete our knowledge on the still unknown redshifts for many BL Lacs~\citep{Goldoni2021}.  The~Imaging X-ray Polarimetry Explorer (IXPE) launched on 2021 December 9 will observe several AGN and blazars during its first years and already provided spectropolarimetric constraints on the radiogalaxy Cen A, suggesting that the X-rays from its core are due to Compton scattering of low-energy photons by non-thermal electrons accelerated in regions with highly disordered magnetic fields within a few parsecs of the black hole, possibly around the innermost jet~\citep{Ehlert2022}. In~blazars, X-rays explore either the high-frequency part of the synchrotron bump, or~the low-frequency part of the (leptonic or hadronic) high-energy bump, and~X-ray polarization will be a strong diagnostic of the different radiation and acceleration models. IXPE and other upcoming X-ray polarization missions, also combined with MWL data, should shed new light on many fundamental questions of blazar physics, such as the magnetic field properties and the nature of the dominant emission processes~\citep{Peirson2022}, the~relevance of multiple emission zone models in turbulent jets~\citep{Peirson2019,Marscher2021}, and~the relative importance of shocks, turbulence and relativistic magnetic reconnection to accelerate particles during HBL flares~\citep{Zhang2021,DiGesu2022}. 

Indeed IR and X-ray data have already drastically modified our standard view of the so-called ``dusty tori'' which now appear clumpy and dynamic across various spatial scales, and~often present a polar component in addition to the equatorial toroidal structure~\citep{RamosAlmeida2017}. IR interferometry of many local AGN has shown two nuclear components on scales of 0.1 to 10 pc, possibly described as an equatorial inflowing disk and a polar-extended feature due to a dusty outflowing wind~\citep{Hoenig2017}. If~applied to FSRQ, such results would alter the properties of the expected ambient photon field around and between the broad and narrow line regions and would have a significant impact on $\gamma$-$\gamma$ absorption and external inverse-Compton emission at the location of the nuclear jet.   

The next generation of gamma-ray space telescopes will be crucial for simultaneously accessing the entire gamma-ray domain. In~particular, projects such as Amego and Amego-X, the~All-sky Medium Energy Gamma-ray Observatory, should fill the ``MeV gap'' in sensitivity~\citep{Engel2022}. Gamma-ray polarimetry in principle has the ability to distinguish between different radiation mechanisms and to probe hadronic signatures in blazars. The~detection of a high MeV polarization would be a direct indication that the proton synchrotron dominates the high energy bump of the~source

Other MWL missions, such as the Transient High Energy Sky and Early Universe Surveyor (THESEUS) with two high-energy instruments covering X-rays and soft gamma-rays and one IR telescope on board, selected for an assessment study with a possible launch in the 2030s, will be especially important to explore the time domain of AGN and blazars~\citep{Mereghetti2021}. Due to its large field of view, THESEUS should for instance be able to catch and follow transient phenomena from a large population of AGN, discover new cases of  Quasi-Periodic Eruptions (QPE), and~study the peculiar ``changing state AGN'' which evolve from one AGN type to another one over various timescales. It should gather very accurate time lags between various high-energy bands, and~cover the full X-ray duty cycle of HBL flares, directly probing particle acceleration and radiation~processes.  

\subsection{Synergy with Radio VLBI and Absolute Sub-Mas~Astrometry}

\begin{figure}[H]
\includegraphics[width=12.5 cm]{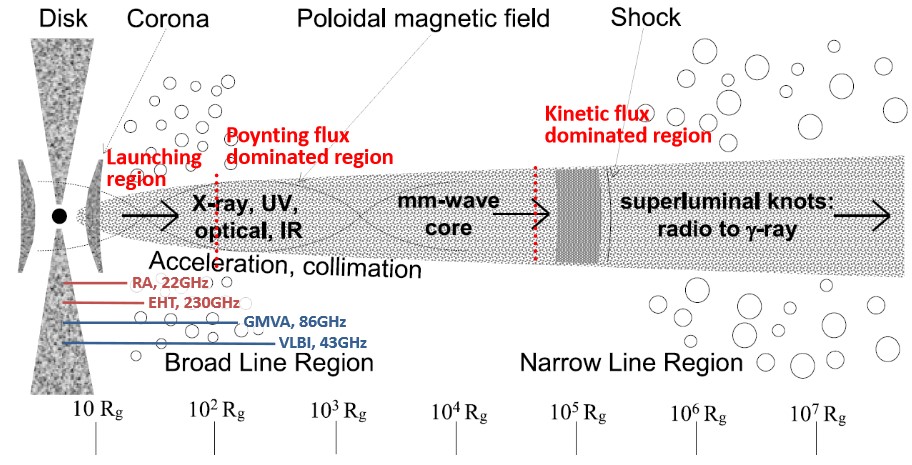}
\caption{Innermost regions of an AGN: standard sketch of a relativistic jet, showing the launching, collimation, acceleration and beginning of propagation zones as a function of the BH gravitational radius $R_g$. A~typical resolution is given here for an AGN at low redshift, for~four global and space (RadioAstron) VLBI networks at different frequencies, illustrating the possibility to well explore the jet collimation and propagation at the multi-parsec scale, and~to probe even into the jet launching region far below the parsec scale, depending on the opacity in the jet and surroundings. Taken from~\cite{Venturi2020}.}
\label{fig:Sketch-jet-VLBI}
\end{figure}

As already illustrated in Section~\ref{subsec:nuclearjets}, the~synergy between the very high angular resolution provided by radio VLBI and the very high temporal resolution explored by IACT at VHE appears very promising, already now (see Figure~\ref{fig:Sketch-jet-VLBI}) but also with the first operations of the SKA, CTA, and~other facilities expected for this decade, together with the concept of Global VLBI Alliance under development, which will change the landscape for the study of the non-thermal universe~\cite{Venturi2020}. VLBI networks will have the power to resolve blazars down to their inner core-jet region. Fast and multi-frequency VLBI imaging in total and polarized light of the vicinity of supermassive black holes and nuclear jets for a sample of gamma-ray emitting AGN will explore the detailed properties and the evolution of the radio-emitting plasma and of the magnetic field, and~should clarify the jet-launching, particle acceleration and energy release mechanisms. Coordinated with VHE monitoring, this might be the unique way to firmly and accurately identify the location of the VHE flare emitting zones, and~to understand the respective role of the black hole magnetosphere and nuclear jet in the generation of quiescent and fast-varying gamma rays. The~cases of binary black hole VHE blazars and of gravitationally lensed ones, although~more complex, might in addition offer the opportunity to observe the same relativistic jet from different viewing angles and configurations, providing a first insight into the three-dimensional properties of the~phenomena.

The International Celestial Reference Frame (ICRF3) and Gaia Data Releases now provide absolute sub-milliarcsecond astrometry at three radio frequencies and in the optical range with a precision of the order of 0.1~mas. 
Positional differences for more than 3500~AGN cross-identified in ICRF3 and EDR3 reveal some significant offsets between the radio and optical centroids at the mas level. A~part of the optical centroids appear coincident with VLBI knots, and~a few of them even show high linear polarization~\cite{Kovalev2020,Lambert2021}. Such results suggest that some of the Gaia centroids from blazars are dominated by a non-thermal synchrotron optical emission from an active zone along the nuclear jet, and~could in such a case be a direct signature of the location of the gamma-ray emitting zone~\cite{Lambert2022}. This is particularly promising for BL Lacs where other contributions to the optical emission, such as radiation from the accretion disk, are expected to be very low. Progress in this area will require coordination between VHE monitoring and sub-mas astrometry surveys in the radio and optical~domains.   

\subsection{The Link with Neutrinos and~UHECRs}
\label{subsec:neutrinos}

Due to their deflection in intergalactic and Galactic magnetic fields, it is notoriously difficult to establish the sources of ultra-high-energy cosmic rays (UHECRs) as presently observed with cosmic ray detectors such as the Telescope Array and the Pierre AUGER Observatory. Spatial correlation between the arrival directions of UHECR and various populations of VHE cosmic sources such as AGN have been analyzed, but~no significant one was found up to now~\cite{Coleman2022}. More specific answers are expected soon with the upgraded detectors, and~later with new-generation experiments such as POEMMA, GRAND and GCOS. It is also interesting to note that, since AGN jets accelerate electrons, protons and ions, very long term jet variability could be related to features of the UHECR spectrum~\citep{Matthews2021}. For~the time being, only the detection by IceCube, or~by new projects like the KM3NeT in the years to come, of~high-energy neutrinos from astrophysical objects permits to directly infer the acceleration of (U)HECRs in well identified cosmic sources. Lepto-hadronic emission models then provide a direct link between a population of cosmic rays and the expected MWL and neutrino emission for a given~scenario.

Given the low detection rate of astrophysical neutrinos and the limited angular resolution of their arrival directions, temporal correlations with variable emitters help to significantly improve the identification of their sources. The~recent first detection of a high-energy neutrino from the blazar TXS\,0506+056 in a flaring state~\cite{IceCube2018} provided thus an important opportunity to compare different emission scenarios to the MWL and multi-messenger data. A~large number of lepto-hadronic scenarios have been confronted to this data set\endnote{cf. 
 \url{http://tevcat.uchicago.edu/?mode=1\&showsrc=309} (accessed on 28 October 2022).}; a short review is given in~\cite{Cerruti2020}. The~scenarios that most convincingly describe the observed MWL and neutrino emission during the flaring state imply a combination of an SSC model and a hadronic component that requires in addition external photon fields, from~the surrounding jet or emission regions in the AGN, to~reach the estimated neutrino flux. The~subsequent detection of a neutrino flare from the same source, which was not accompanied by any significant flux increase in the MWL emission~\cite{IceCube2018b}, presents clearly a challenge for single-zone lepto-hadronic models, and~seems to indicate that neutrinos and gamma-rays may well be produced in separate emission~regions.

Statistical studies suggest as well that jetted AGN with strong Doppler boosting could be important sources of neutrinos, as~shown for instance by the fact that the sub-group of AGN associated with the highest energy neutrinos (E > 200 TeV) appears more core-dominated than the whole complete VLBI-flux-density limited sample analyzed in ~\citep{Plavin2020}. It has also been pointed out that in some blazars that may coincide with neutrinos recorded by IceCube, the~gamma-ray emission is temporarily suppressed during efficient neutrino production due to increased 
$\gamma$-$\gamma$ opacity~\cite{Kun2021}. If~its existence were confirmed, such a particular class of blazars could contribute to the isotropic cosmic flux of HE neutrinos discovered by IceCube in 2013~\cite{IceCube2020}, the~origin of which is still~unknown.  

\subsection{Perspective with Gravitational~Waves}

Our knowledge of the gravitational wave (GW) universe will improve dramatically over the next few decades. Several events directly related to the massive BH of blazars and AGN, such as merger, capture of compact stars, spiralling phase, should be explored by Pulsar Timing Arrays and future gravitational wave space interferometers like LISA, whose launch is expected in 2034. As~already observed in GRB and in TDE, the~spiralling phases can lead to the formation of relativistic jets from which an emission of gamma-rays is expected and could be detected for ``not too distant'' transient jets directed towards the solar system. CTA's wide field of view would be well suited to search for such gamma counterparts of GW~events.

Currently, the~origin and formation sites of the population of binary black hole (BBH) mergers detected by LIGO-Virgo remains uncertain as electromagnetic counterparts are very weak and difficult to detect in the absence of a dense gaseous environment. However there are theoretical arguments in favor of a possible AGN-disk origin for some BBH events. It could be the case of GW190521 for which an optical AGN flare was suggested as a possible electromagnetic counterpart (although debated)~\citep{Graham2020}. Standard AGN accretion disks are conducive to stellar mass BBH mergers and can lead to hierarchical formation of black holes, providing an explanation for the large BH masses currently observed. Such environment also induce an excess of eccentric mergers like GW190521~\citep{Samsing2022}, and~could produce significant electromagnetic counterpart due to the dense ambient plasma surrounding the merger. Indeed, non-zero eccentricity and non-zero spin--orbit tilt directly found in gravitational data could characterize the mergers occurring in AGN disks, in~contrast to isolated mergers or mergers in stellar clusters. This might be difficult to achieve with LIGO-Virgo but will be possible with the Einstein telescope project. Prompt optical-UV counterpart due to the formation of shocks has been predicted from such stellar mass mergers, when the sphere of influence of the binary system is larger than the disk thickness, or~depending on the disk opacity~\citep{McKernan2022}. Transient broad band emission from radio to high energies, as~well as HE neutrinos and cosmic rays, can also be expected from 10 to 100 solar mass mergers in AGN disks during episodes where the BBH accrete sufficient gas at high rate, exceeding the Eddington limit and generating fast outflows and transient relativistic jets through the formation of mini-disks~\citep{Murase2016,Bartos2017}. Gamma-rays could be detectable with present and future instruments in the case of energetic outflows, even before the merging event. Such MWL and MM events would correspond to a very specific type of blazar flares which have not yet been much considered, and~would provide completely new information on AGN formation and accretion disks~properties.

\section{Conclusions and~Outlook}
\label{sec:conclusion}

Beyond blazar physics, blazar data and modelling at VHE are also mandatory for several questions of fundamental physics and cosmology, gamma-ray emitting blazars being commonly used as beacons to explore their line of sight and probe the nature of space-time through the analysis of the propagation of their VHE photons on cosmic scales. Any progress in the understanding and description of the intrinsic gamma-ray spectra and variability of blazars will significantly impact all results on EBL, intergalactic magnetic fields (IGMF), Lorentz invariance violation (LIV) studies, and~axion-like particles searches. A~better knowledge on blazar behavior would also usefully contribute improving the accuracy of celestial reference frames, which use a majority of blazars as cosmic reference points and are essential for many activities such as geophysics, GPS, space navigation. Blazar research clearly has a multidisciplinary interest. As~illustrated in this review, fundamental difficulty in consolidating any emission modeling is the current absence of a general baseline scenario of blazars and active galactic nuclei sufficiently developed to consistently  include the new information collected in recent years. This should be overcome in the coming decade where a wealth of high quality data is expected from several new MWL and MM infrastructures, with~the advances in computing power and theoretical~models.

\vspace{6pt}

\conflictsofinterest{The authors declare no conflict of interest.}


\begin{adjustwidth}{-\extralength}{0cm}
\printendnotes[custom] 

\reftitle{References}




\end{adjustwidth}

\end{document}